# Nano-Mechanical Behavior of MnTiO$_3$ Ceramics

R. K. Maurya[1-2]* and Davinder Singh[2]

[1]School of Basic Sciences, Indian Institute of Technology Mandi, Kamand, Himachal Pradesh-175005, India.
[2]Department of Physics, IISER Bhopal, Bhopal-462066, India.
[2]School of Engineering, Indian Institute of Technology Mandi, Kamand, Himachal Pradesh-175005, India

*rkumar.0121@gmail.com      (2 is the present address of author)

**Abstract:** Here we investigate the mechanical behavior of the MnTiO$_3$ ceramic. Nanoindentation tests in a manganese titanate polycrystalline sample were performed using a Berkovich indenter with a tip radius of 150 nm. On applying the load, it shows the elastic and plastic behavior. The work done during the elastic and plastic processes are 16757.78 μN.nm and 10045.72 μN.nm, respectively. The estimated values of reduced modulus and hardness are 158±8 GPa and 10.5±2GPa, respectively. This suggests that this compound exhibits a significant plastic character.

## 1. Introduction

Ceramics are very important candidates of the material science due to their fundamental physics and application point of view. Ceramic materials are very hard at ambient temperature and pressure due to the strong ionic or covalent bonding existing in it. Metal titanates with versatile performances have attracted rapidly growing interest over the past few decades, mainly because of their unique properties and important technological applications. MnTiO$_3$ is an important material having promising application in high density data storage, optical, solar cells, humidity and magnetic field sensors, dielectrics and photocatalytic ceramics[1-3]. It is a low cost, abundant and highly stable naturally occurring ceramic. MnTiO$_3$ ceramic has also been studied broadly due to their exotic physical properties. MnTiO$_3$ is one of the members of the ilmenite family ATiO$_3$ (A=Fe, Mn, Co, Ni). These ilmenite materials have become very popular candidates due their structural, electronic, optical and magnetoelectric properties[4,5]. It is a multiferroic material which shows the multiferroicity only in one direction (along c-axis)[5].

Nanoindentation is widely used technique for measuring the elastic modulus, E, and the hardness, H, of small volumes of materials and thin film. This technique has also become useful to measure the other mechanical parameters such as hardening exponents[6], residual stress[7] and creep parameters[8]. However, this technique provides more fundamental inquiries of the material science. In a standard nanoindentation test the elastic and plastic material response is evaluated. Previously limited reports are available on mechanilcal behavior of metal titanates. *Beirau et al.* provided new insights on effect of radiation elastic modulus and hardness of naturally occurring titanates[9].

The motivation of this work is to address the mechanical reliability of MnTiO$_3$ ceramics for potential device applications. In this work, we report successful one step synthesis of manganese titanate and their nanomechanical behavior.

## 2. Experiment

The highly crystalline sample of MnTiO$_3$ was prepared using conventional solid state route. The starting materials of MnCO$_3$ and TiO$_2$ were mixed and ground under ethanol using mortar and pestle. The mixture was heated at 1200˚C for 24 hours in air. The single phase sample was obtained after several intermediate grindings. The sample was characterized using x-ray diffraction, Raman spectroscopy, Scanning electron microscopy (SEM) and nanoidentation measurements. The x-ray diffraction (XRD) experiment confirms it as single phase. The room temperature x-ray diffraction measurement was performed using 9kW rotating anode x-ray diffractometer with Kα radiation. The x-ray diffraction pattern was analyzed with Rietveld profile refinement method using FullProf software[10]. The room temperature Raman spectra were collected using a LabRam (Horiba Scientific Inc.) spectrometer in a back scattering geometry. The microstructural characterization was done using scanning electron microscopy (FEI Nova Nano SEM-450). The nanoidentation tests were performed using TI 950 TriboIdenter (Hysitron, USA) with three sided diamond Berkovich tip.

## 3. Results and Discussion

The three dimensional (3D) crystal structure of MnTiO$_3$ is shown in Fig.1(a). In the 3D crystal structure of MnTiO$_3$, the same type of octahedra (MnO$_6$ or TiO$_6$ ) are connected to each other by edge sharing and opposite type of octahedra (MnO$_6$ and TiO$_6$) are connected to each other face sharing. In this compound, Mn$^{2+}$ and Ti$^{4+}$ ionic layers are arranged alternatively along the hexagonal c-axis. The cations form a distorted honeycomb lattice along the c-axis due to the electrostatic repulsion between them. The magnetism in this compound arises due the partially filled d-orbitals in Mn$^{2+}$ ions.

Within the layer (intra-layer), two Mn spins interact to each other through O ion via superexchange interaction and between the layer (inter-layer) through two O ions via super-superexchange interaction[11]. The intra-layer interactions are stronger than the inter-layer interactions.

The Rietveld refinement of x-ray diffraction pattern of MnTiO3 collected at 300K is shown in Fig. 1(b). It stabilizes in the centro-symmetric space group $R\bar{3}$. The XRD pattern does not show any additional reflection other than the MnTiO$_3$ phase. The room temperature lattice parameters obtained from the Rietveld refinement are a = b= 5.136(9) Å, c=14.278(4) Å. The lattice parameters are in agreement with our previous reports[11]. The goodness of fit parameter, S is ~ 1.33.

Fig.2 shows the Raman spectra of MnTiO$_3$ collected at the room temperature. Ten Raman active bands were observed during the measurements which are in line with the reported results[12]. The frequency shift of these peaks at room temperature are 164, 202, 236, 263, 336, 360, 479, 609, 684 cm$^{-1}$, respectively.

The scanning electron microscopy image recorded on the pellet of this material is shown in Fig. 3. The physical properties of the materials get modified due to the variation in the grain size. The SEM micrographs clearly show that the material is homogeneous. No voids have been observed in the sample. It is observed that the grains are almost of the same size and different in shape. The different shapes of the grains indicate that the material is polycrystalline in nature. The average grain size is found to be in the range of 6-15μm. On the basis of SEM micrographs, the average grain size observed by *T. Acharya et al*[13]. is in the range of 3-10μm which is less than we observed. This is due to the less sintering temperature and time used by them. Usually, on increasing the sintering temperature and time, the grain size increases which support our results also.

Fig.4 shows the Nanoindentation of $MnTiO_3$ during the loading and unloading of indenter into the sample. The inset figure shows the scanning probe microscopy (SPM) image of the impression. During the loading process the atoms get pressed while in the unloading process the atoms recover partial portion of total displacement.

In a typical nanoindentation test applied test load, P and indentation depth, h of the indenter is continuously measured. Here, load was increased from 0 to 1200 µN in 5 s, held there for 2 s and decreased back to 0 µN in 5 seconds. We observed the ceramic elastically recovered almost a quarter of the maximum penetration depth, hmax at max applied load. This recovered depth is called contact depth, hc measured from intersection of slope of unloading curve with x axis, is required to estimate area, Ac of the residual impression[14]. Total indention work done is 26803.50 µN.nm. The work done during the elastic and plastic processes are 16757.78 µN.nm and 10045.72 µN.nm., respectively. The work done in the elastic process is more than the plastic process. This suggests that the $MnTiO_3$ exhibits plastic character too. The plasticity has also been observed in other ceramics[15]. Stiffness, S of the elastic contact between indenter and specimen during unloading is estimated from the slope of the unloading curve and is related to the reduced modulus, Er as[14]

$$Er = \frac{\sqrt{2}}{\pi} \frac{S}{\sqrt{A}}$$

The estimated values of reduced modulus and hardness of the materials was also of subtle interest. Further the expression for the hardness, H is given as

$$H = \frac{P_{max}}{A}$$

Here P is the applied load and A is the indent area. The extracted values of reduced modulus (Er) and hardness (H) are 158±8 GPa and 10.5±2 GPa, respectively.

## 4. Conclusion

Our results show that the $MnTiO_3$ stabilizes in the hexagonal crystal structure with the space group $R\bar{3}$. The x-ray diffraction and Raman spectroscopy results confirm it as single phase. The SEM result confirms the homogeneity and polycrystalline nature of the material. The average grain size observed by the SEM measurements is in the range of 6-15µm. During the nanoindentation experiments, we have observed that the work done in the elastic process is more than the plastic process. This suggests that $MnTiO_3$ reveals a plastic character in addition to the elastic character.

**Acknowledgement**

Authors thank the Advanced Material Research Centre (AMRC), IIT Mandi for providing the characterization facilities. Authors are also thankful to Dr. Viswanath Balakrishnan and Dr. Bindu Radhamany for the fruitful discussion.

**References**

1. M. Shaterian, M. Barati, K. Ozaee and M. Enhessari, J. Ind. Eng. Chem. **20**, 3646 (2014).
2. H. Y. He, J. F. Huang, L. Y. Cao and J. P. Wu, Sensor Actuat B- Chem **132**, 5-8 (2008).
3. K. F. Wang, J. M. Liu, Z. F. Ren, Adv. Phys. **58**, 321 (2009).
4. R. K. Maurya, Priyamedha Sharma, Ashutosh Patel and R. Bindu, Europhys. Lett. **119**, 37001 (2017).
5. N. Mufti, G.R. Blake, M. Mostovoy, S. Riyadi, A.A. Nugrohu and T.T.M Palstra, Phys. Rev. B **83**, 104416 (2011).
6. M. Dao, N. Chollacoop, K. J. Van Bliet, T. A. Venkatesh and S. Suresh, Acta. Mater. **49**, 3899-3918 (2001).
7. A. C. Fischer-Cripps, Mater. Sci. Eng. A **385**, 74-82 (2004).
8. S. Suresh and A. E. Giannakopoulos, Acta. Mater. **46**, 5755-5767 (1998).
9. T. Beirau, W. D. Nix, R. C. Ewing, G. A. Schneider, L. A. Groat and U. Bismayer **101**, 399-406 (2016).
10. J. R. Carvajal, FULLPROF, a Rietveld refinement and pattern matching analysis program, Laboratoire Leon Brillouin CEA- CNRS, France, 2000; 7- H.M. Rietveld, Acta Crystallogr. **22**, 151(1967); J.Appl. Cryst.**2**, 65 (1969); A.W. Hewat, Haewell Report No.73/239 (1973), ILL Report No. 74/H62S (1974); G. Malmros and J.O. Thomas, J. Appl.Cryst. **10**, 7 (1977); C.P. Khattak and D.E. Cox ibid.**10**, 405 (1977).
11. R. K. Maurya, Navneen Singh, S. K. Pandey and R. Bindu, Europhys. Lett. **110**, 27007 (2015).
12. J. Ko, N. E. Brown, A. Navrotsky, C. T. Prewitt and T. Gasparik, Phys. Chem. Minerals **16**, 727-733 (1989).
13. Truptimayee Acharya and R. N. P. Choudhary, Appl. Phys. A. 121, 707-714 (2015).
14. C. Oliver and M. Pharr, J. Mater. Res.**7**, 1564-1583 (1992).
15. X. Xu, Y. Wang, A. Guo, H. Geng, S. Ren, X. Tao and J. Liu, Int. J. Plast. **79**, 314-327 (2016).

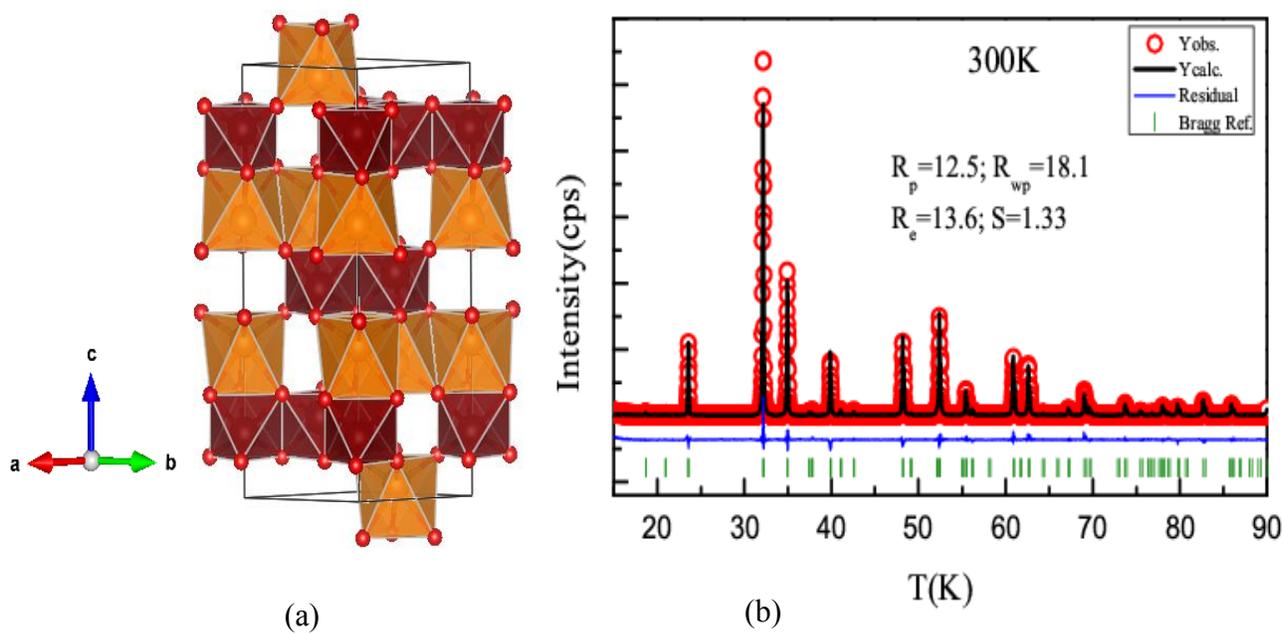

Fig.1: (a) Crystal structure of MnTiO$_3$ in which Mn$^{2+}$ (yellow) and Ti$^{4+}$ (wine) are octahedrally coordinated with the O$^{-2}$ (red) atoms, (b) X-ray diffraction pattern of MnTiO$_3$ at 300K with the different profile parameters. The open circles and solid lines correspond to the observed and the calculated patterns, respectively. The vertical bars represent the position of the Bragg reflections and blue line shows the difference between the observed and calculated intensities.

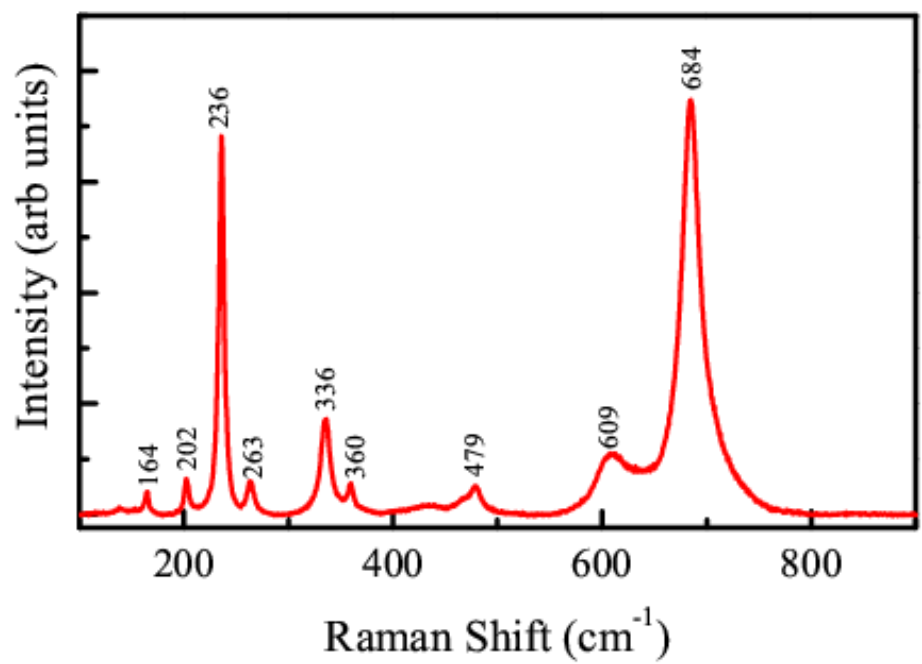

Fig.2: Raman spectra of MnTiO$_3$ at room temperature, collected on the pellet.

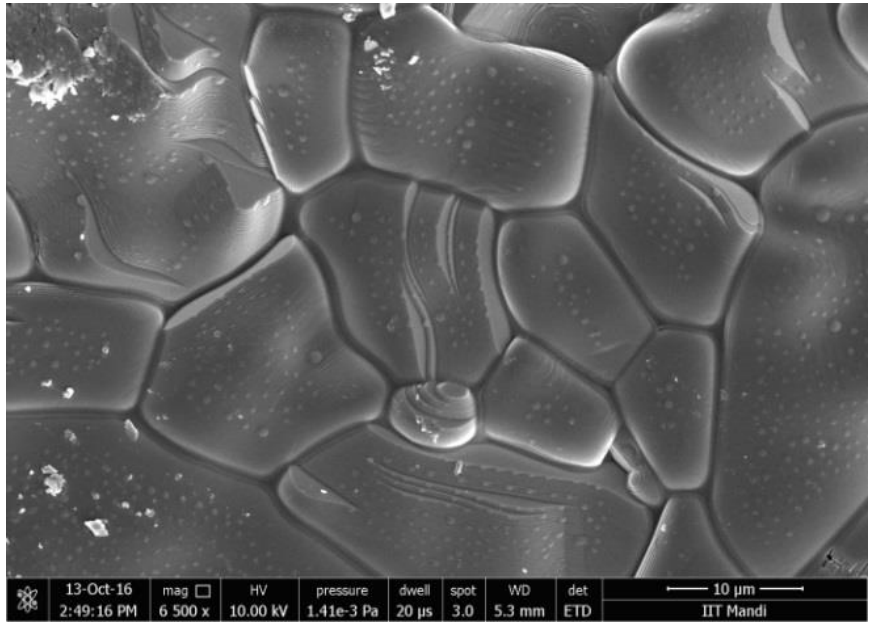

Fig.3: SEM micrograph of MnTiO$_3$ recorded on the pellet at room temperature.

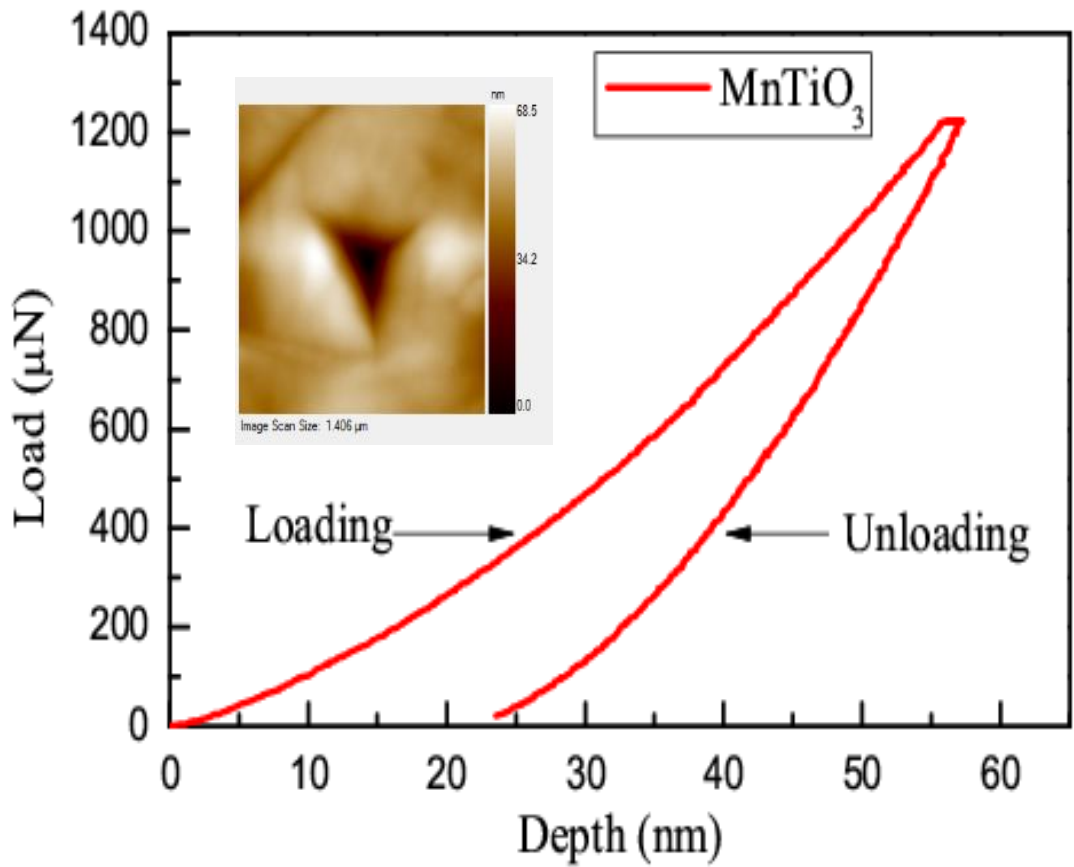

Fig. 4: Loading unloading curve of $MnTiO_3$ during applied load of 1200 μN. The inset shows the SPM image of the sample.